# Partition functions for two-dimensional Ising models using Generalised Hypergeometric series and Chebyshev polynomials

M V Sangaranarayanan

Department of Chemistry

Indian Institute of Technology-Madras Chennai-600036

India

E-mail: sangara@iitm.ac.in

**Abstract:**

The zero-field partition function of two-dimensional nearest neighbour Ising models of square lattices is derived in terms of the generalized hypergeometric series $_4F_3\left(1,1,\frac{3}{2},\frac{3}{2};2,2,2;\kappa^2\right)$ by evaluating the integral in the Onsager's exact solution. An approximate eqn for the partition function in terms of Chebyshev polynomials is also provided. The spontaneous magnetization equations derived using Bragg-Williams approximation and Onsager's exact solution are also expressed using the simple Gauss hypergeometric series $_2F_1(a, b, c, z)$.

## 1. Introduction

The analysis of critical properties of Ising and lattice gas models continues to be a frontier area of research in diverse branches of physics [1]. In particular, the two-dimensional Ising models have evoked considerable interest in diverse contexts [2-4] and Onsager's exact solution at zero magnetic field has yielded an explicit expression for the critical temperatures apart from predicting the logarithmic singularity of the specific heat [5]. The analysis of Yang [6] deserves mention in this context on account of a simpler expression for the spontaneous magnetization. Other notable methodologies propounded for the study of Ising models include: Bragg-Williams zeroth order approximation [7], Bethe quasi-chemical approach [8], series expansions [9], renormalization group *ansatz* [10], scaling hypothesis [11], graph theoretical procedures [12] etc.

The zero field partition function and the internal energy deduced therefrom by Onsager [5] are analytic functions, while the specfic heat exhibits a logarithmic singularity at the critical temperature. Hence, it is of interest to analyze the partition function and related quantities using special functions of mathematical physics so that new insights may consequently emerge.

The objectives of the present communications are as follows: (i) to report the partition function arising from Onsager's exact solution of two-dimensional Ising models in terms of the generalized hypergeometric series $_4F_3\left(1,1,\frac{3}{2},\frac{3}{2};2,2,2;\kappa^2\right)$ using a simple integration procedure (ii) to provide an approximate expression for the partition function with the help of Chebyshev polynomials and (iii) to formulate the spontaneous magnetization of Bragg-Williams approximation (BWA) and Onsager's exact solution using Gauss hypergeometric series $_2F_1(a,b,c;x)$.

## 2. Onsager's exact solution for the partition function for square lattices

The two-dimensional nearest neighbour Ising Hamiltonian is given by [4]

$$H_T = -J\left[\sum_{ij}(\sigma_i\sigma_j\sigma_i\sigma_{j+1} + \sigma_i\sigma_j\sigma_{i+1},\sigma_j)\right] - H\sum_i \sigma_j \qquad (1)$$

where $J$ denotes the interaction energy, $H$ being the external magnetic field. The spin variable takes either of the two values $+1$ or $-1$. The subscripts $i$ and $j$ denote the row and column indices. The partition function $Q$ for the above Hamiltonian has been deduced for a square lattice of $N$ sites as [5]:

$$\left(\frac{1}{N}\right)\log Q = \log[\sqrt{2}\cosh(2K)] + \left(\frac{1}{\pi}\right)\int_0^{\pi/2} d\phi \log\left[1 + \sqrt{1-\kappa^2\sin^2(\phi)}\right] \qquad (2)$$

where

$$\kappa = \frac{2\sinh(2K)}{\cosh^2(2K)} \qquad (3)$$

with $K = J/kT$. Although the numerical evaluation of the intergal in eqn (2) is straight forward, it is of interest to verify whether any 'closed form' expressions (either exact or approximate) for the partition function can be obtained. It has also been stated that "the integral can not be expressed in closed form" [5].

### 2.1 Partition function in terms of generalised hypergeometric series

The integral appearing in eqn (2) has been earlier evaluated [12] by identifying it as belonging to Mahler measure which itself is related to Gauss hypergeometric functions and by comparing the coefficients in the 'formal power series' with those from Gauss hypergeometric series. The resulting partition function is shown to be

$$\frac{1}{N}\log Q = \log 2 + \log[\cosh(2K)] - \kappa_1{}^2\,{}_4F_3\left(1,1,\frac{3}{2},\frac{3}{2};2,2,2;16\kappa_1{}^2\right) \quad (4)$$

with the definition of $\kappa_1$ as

$$\kappa_1 = \frac{\tanh(2K)}{2\text{sech}(2K)} \quad (5)$$

Here, we propose a simple method for evaluating the integral. For this purpose, let us denote the integral of eqn (2) as $I_1$ and define another auxiliary integral $I_2$ as

$$I_2 = \int_0^{\frac{\pi}{2}} \log\left(1 - \sqrt{1-\kappa^2\sin^2\phi}\right)d\phi \quad (6)$$

Obviously,

$$I_1 + I_2 = \int_0^{\pi/2} \log(\kappa^2 \sin^2\phi)d\phi = \frac{\pi}{2}\log\left(\frac{\kappa^2}{4}\right) \quad (7)$$

However

$$I_1 - I_2 = \int_0^{\pi/2} 2\tanh^{-1}\left(1 - \sqrt{1-\kappa^2\sin^2\phi}\right)d\phi \quad (8)$$

The evaluation of the above integral is more involved and can be carried out [15] as shown in the Appendix A, yielding the integral $I_1$ as

$$I_1 = \frac{\pi}{2}\log 2 - \frac{\pi}{16}\kappa^2\,{}_4F_3\left(1,1,\frac{3}{2},\frac{3}{2};2,2,2,\kappa^2\right) \quad (9)$$

Thus the partition function becomes

$$\frac{1}{N}\log Q = \log 2 + \log[\cosh(2K)] - \frac{1}{16}\kappa^2\,{}_4F_3\left(1,1,\frac{3}{2},\frac{3}{2},2,2,2,\kappa^2\right) \quad (10)$$

with $\kappa$ being given by eqn (3).

Although the above eqn is similar to the one derived earlier [12] with an alternate definition of $\kappa$ (cf. eqn (5)), eqn (10) has resulted from a simple procedure for evaluating the integral [14,15] (Appendix ).

### 2.2 Partition function interms of Chebyshev polynomials

It is of interest to anlayze whether other methods (even if approximate) of writing the partition function exists. It is well known [16] that several elementary functions can be approximated using Chebyshev polynomials. For logarithmic series, Glenshaw [17] has provided the first few Chebyshev coefficients as follows:

$$\log(1+x) = \sum_{n=0}^{\infty} A_n T_n^*(x) \quad (11)$$

with $T_n^*(x) = \cos(n\theta), \quad \cos\theta = 2x - 1$

The first few coefficients of $\{A_n\}$ are given in Table 1.

**Table 1: The first few coefficents of Chebyshev polynomial of eqn ( 11).**

| $n$ | $A_n$ |
|---|---|
| 0 | 0.37645 |
| 1 | 0.34315 |
| 2 | −0.02944 |
| 3 | 0.00337 |

Consequently,

$$\log(1+x) \approx 0.37645 + 0.34315(2x-1) - 0.02944(8x^2 - 8x + 1)$$
$$+ 0.00337(32x^3 - 48x^2 + 18x - 1)$$
$$\approx 0.00049 + 0.98248x - 0.39728x^2 + 0.10784x^3 \quad (12)$$

It has been shown by Glenshaw [17] that the above eqn has a maximum error of 0.00053 for $x$ ranging from 0 to 1 and interestingly $\kappa$ also lies in this range only. By identifying $x$ as $\sqrt{1-\kappa^2\sin^2\phi}$ and carrying out the integration term by term, the integral $I_1$ follows as

$$I_1 \approx \frac{\pi}{2}\left[-0.39679 + 0.19864\kappa^2 + {}_2F_1\left(-\frac{1}{2},\frac{1}{2},1,\kappa^2\right)(1.12626 - 0.07189\kappa^2) + {}_2F_1\left(\frac{1}{2},\frac{1}{2},1,\kappa^2\right)(0.03594)(\kappa^2 - 1)\right] \quad (13)$$

In terms of the elliptic integrals, we can write in an equivalent manner,

$${}_2F_1\left(-\frac{1}{2},\frac{1}{2},1,\kappa^2\right) = \frac{2}{\pi}E(\kappa) \quad \text{and} \quad {}_2F_1\left(\frac{1}{2},\frac{1}{2},1,\kappa^2\right) = \frac{2}{\pi}K(\kappa) \quad (14)$$

Hence the integral can be approximated as

$$I_1 \approx \frac{\pi}{2}\left[-0.39679 + 0.19864\kappa^2 + \frac{2}{\pi}E(\kappa)(1.2626 - 0.7189\kappa^2) + \frac{2}{\pi}K(\kappa)(0.03594)(\kappa^2 - 1)\right] \quad (15)$$

where $K(\kappa)$ and $E(\kappa)$ are respectively the complete elliptic integrals of the first and second kind. Table 2 provides the estimates from eqn ( 15) along with the values from direct numerical integration. It is inferred that the values are in excellent agreement with the numerical integration results.

Alternately, $I_1$ can be written as

$$I_1 = \int_0^{\pi/2} d\phi \log[1 + \sqrt{1 - \kappa^2 \sin^2(\phi)}] = \int_0^{\frac{\pi}{2}} d\phi \sum_{n=0}^{\infty} A_n T_n^*(\sqrt{1 - \kappa^2 \sin^2(\phi)}) \quad (16)$$

with the first few Chebyshev coeffcients { $A_n$} shown in Table 1. Eqn (2) can be rewritten

$$\left(\frac{1}{N}\right)\log Q = \log[\sqrt{2}\cosh(2K)] + \left(\frac{1}{\pi}\right)\int_0^{\pi/2} d\phi \sum_{n=0}^{3} A_n T_n^*(\sqrt{1 - \kappa^2 \sin^2(\phi)}) \quad (17)$$

Hence

$$\frac{1}{N}\log Q = \log 2 + \log[\cosh(2K)] + \frac{1}{2}\Big[-0.39679 + 0.19864\kappa^2 + \frac{2}{\pi}E(\kappa)(1.2626 - 0.7189\kappa^2) + \frac{2}{\pi}K(\kappa)(0.03594)(\kappa^2 - 1)\Big] \quad (18)$$

**Table-2- A few values of the integral $I_1$ estimated using eqn ( 15 ) and comparison with the numerical integration result**

| $\kappa^2$ | value from eqn ( 15 ) | Numerical integration value |
| --- | --- | --- |
| 0 | 1.08939 | 1,08879 |
| 0.1 | 1.06873 | 1.06858 |
| 0.2 | 1.04703 | 1.04708 |
| 0.3 | 1.02389 | 1.02408 |
| 0.4 | 0.99903 | 0.99927 |
| 0.5 | 0.97200 | 0.97223 |
| 0.6 | 0.94223 | 0.94236 |
| 0.7 | 0.90866 | 0.90870 |
| 0.8 | 0.86963 | 0.86951 |
| 0.9 | 0.82098 | 0.82096 |
| 1.0 | 0.74311 | 0.74313 |

Eqn (18) although approximate is entirely new and the presence of elliptic integrals $K(\kappa)$ and $E(\kappa)$ in the partition function itself brings about a consistency with the internal energy and specific heat equations (eqns 116 and 117 of Onsager [ 5]). For anisotropic Ising models too, a double hypergeometric function in two variables has been deduced recently [18].

### 3. Spontaneous Magnetization using Gauss hypergeometric series

The exact eqn for the spontaneous magnetization given by Yang [6] is as follows:

$$M_{ons} = \left[1 - \left\{\sinh^{-4}\left(\frac{2J}{kT}\right)\right\}\right]^{1/8} \quad (19)$$

for square lattices and with the critical exponent β as 1/8. On the other hand, the analogous eqn $M_0$ for Bragg-Williams Approximation is

$M_0$ (BWA) $= tanh\ (M_0\ 4J/kT)$ (20)

while the critical exponent β is ½. Interestingly, eqns (19) and (20) may both be written in terms of Gauss hypergeometric functions $_2F_1$ (a,b,c:x) as shown in Table 3.

**Table 3: Equation for spontaneous magnetization in terms of Gauss hypergeometric series**

| Method | Equation for spontaneous magnetization ($M_0$) |
|---|---|
| Bragg-Williams approximation [7] | $4K = {}_2F_1\left(1, \frac{1}{2}, \frac{3}{2}, M_0^2\right) = {}_2F_1(2\beta, \beta, 1+\beta, M_0^{\frac{1}{\beta}})$ with $\beta = 1/2$ |
| Onsager's exact solution [5] | $\sinh(2K) = {}_2F_1\left(\frac{1}{4}, 1, 1, M_0^8\right) = {}_2F_1(2\beta, 1, 1, M_0^{\frac{1}{\beta}})$ with $\beta = \frac{1}{8}$ |

## 4. Results and Discussion

Eqns(10) and (18) are the central results of this study. While eqn (10) is exact and has resulted from the integration in Onsager's exact solution, eqn (18) is an approximation, arising from the use of Chebyshev polynomials for the integration. It is of interst to note the occurrence of Gauss hypergeometric functions of diverse genre for the partition function as well as spontaneous magnetization the analysis of two-dimensional Ising models.

## 5. Summary

The partition function of square lattices pertaining to the two-dimensional nearest neighbour Ising models is written in terms of the generalised hypergeometric series ${}_4F_3\left(1,1,\frac{3}{2},\frac{3}{2};2,2,2;\kappa^2\right)$ by carrying out the integration in Onsager's exact solution using a simple procedure. Another approximate expression for the partition function using Chebyshev polynomials is also reported. The spontaneous magnetizations of Bragg-Williams approximation and Onsgaer's exact result are expressed using Gauss hypergeometric series ${}_2F_1(a, b, c; x)$.

**Acknowledgements**

The financial support by Mathematical Research Impact Centric Support (MATRICS) program of Science and Engineering Research Board, Government of India is gratefully acknowledged.

## Appendix

The steps involved in the evaluation of the integral

$$\int_0^{\frac{\pi}{2}} d\phi \log\left[1 + \sqrt{1 - \kappa^2 \sin^2(\phi)}\right]$$

are shown below[15]:

$$I_1 = \int_0^{\frac{\pi}{2}} \log\left(1 + \sqrt{1 - \kappa^2 \sin^2 \phi}\right) d\phi$$

By defining another integral $I_2$ as

$$I_2 = \int_0^{\frac{\pi}{2}} \log\left(1 - \sqrt{1 - \kappa^2 \sin^2 \phi}\right) d\phi$$

we obtain

$$I_1 + I_2 = \int_0^{\frac{\pi}{2}} \log(\kappa^2 \sin^2 \phi) d\phi = \frac{\pi}{2} \log\left(\frac{\kappa^2}{4}\right)$$

and

$$I_1 - I_2 = \int_0^{\frac{\pi}{2}} 2\tanh^{-1}\left(1 - \sqrt{\kappa^2 \sin^2 \phi}\right) d\phi$$

The integral

$$\int_0^{\frac{\pi}{2}} \tanh^{-1}\left(\sqrt{1 - \kappa^2 \sin^2 \phi}\right) d\phi$$

has been evaluated exactly by a series of substitutions and involving *Beta* functions [ 13]
By substituting $x = \sin(\phi)$, the desired integral becomes

$$\int_0^{\frac{\pi}{2}} d\phi \tanh^{-1}\left(\sqrt{1 - \kappa^2 \sin^2(\phi)}\right) = \int_0^1 dx \frac{\tanh^{-1}\left(\sqrt{1 - \kappa^2 x^2}\right)}{\sqrt{1 - x^2}}$$

Subsequently, another substitution $x = \sqrt{t}$ is made so as to yield

$$\int_0^1 dx \frac{\tanh^{-1}\left(\sqrt{1 - \kappa^2 x^2}\right)}{\sqrt{1 - x^2}} = \int_0^1 dt \frac{\tanh^{-1}\left(\sqrt{1 - \kappa^2 t}\right)}{2\sqrt{t}\sqrt{1 - t}}$$

$$= \int_0^1 dt \frac{\log\left(\frac{1 + \sqrt{1 - \kappa^2 t}}{1 - \sqrt{1 - \kappa^2 t}}\right)}{4\sqrt{t}\sqrt{1 - t}} = \int_0^1 dt \frac{\log\left(\frac{(1 + \sqrt{1 - \kappa^2 t})^2}{\kappa^2 t}\right)}{4\sqrt{t}\sqrt{1 - t}}$$

$$= \int_0^1 dt \, \frac{2\log(1+\sqrt{1-\kappa^2 t}) - \log(\kappa^2 t)}{4\sqrt{t}\sqrt{1-t}}$$

$$= \int_0^1 dt \, \frac{2\log(1+\sqrt{1-\kappa^2 t}) - 2\log(2) + 2\log(2) - \log(\kappa^2) - \log(t)}{4\sqrt{t}\sqrt{1-t}}$$

$$= \int_0^1 dt \, \frac{2\log\left(\frac{1+\sqrt{1-\kappa^2 t}}{2}\right) - \log\left(\frac{\kappa^2}{4}\right) - \log(t)}{4\sqrt{t}\sqrt{1-t}}$$

$$= -\int_0^1 dt \, \frac{\log(\frac{\kappa^2}{4})}{4\sqrt{t}\sqrt{1-t}} - \int_0^1 dt \, \frac{\log(t)}{4\sqrt{t}\sqrt{1-t}} + \int_0^1 dt \, \frac{\log(\frac{1+\sqrt{1-\kappa^2 t}}{2})}{2\sqrt{t}\sqrt{1-t}}$$

$$= \pi\log(2) - \frac{\pi}{4}\log(\kappa^2) + \frac{1}{2}\int_0^1 dt \, \frac{\log(\frac{1+\sqrt{1-\kappa^2 t}}{2})}{\sqrt{t}\sqrt{1-t}}$$

$$= \pi\log(2) - \frac{\pi}{4}\log(\kappa^2) - \frac{1}{8}\int_0^1 dt \, \frac{\kappa^2 t}{\sqrt{t}\sqrt{1-t}}\left[-\frac{4}{\kappa^2 t}\log\left(\frac{1+\sqrt{1-\kappa^2 t}}{2}\right)\right]$$

$$= \pi\log(2) - \frac{\pi}{4}\log(\kappa^2) - \frac{\kappa^2}{8}\int_0^1 dt \, t^{1/2}(1-t)^{-1/2} \, {}_3F_2\left(1,1,\frac{3}{2};2,2;\kappa^2 t\right)$$

$$= \pi\log(2) - \frac{\pi}{4}\log(\kappa^2) - \frac{\kappa^2}{8} B\left(\frac{3}{2},\frac{1}{2}\right) \, {}_4F_3\left(1,1,\frac{3}{2},\frac{3}{2};2,2,2;\kappa^2\right)$$

$$= \pi\log(2) - \frac{\pi}{4}\log(\kappa^2) - \frac{\pi}{16}\kappa^2 \, {}_4F_3\left(1,1,\frac{3}{2},\frac{3}{2};2,2,2;\kappa^2\right)$$

$$\int_0^{\pi/2} 2\tanh^{-1}\left(\sqrt{1-\kappa^2 \sin^2}\right) d\phi = 2\pi\log(2) - \frac{\pi}{2}\log(\kappa^2) - \frac{\pi}{8}\kappa^2 \, {}_4F_3\left(1,1,\frac{3}{2},\frac{3}{2};2,2,2;\kappa^2\right)$$

$$I_1 - I_2 = 2\pi\log 2 - \frac{\pi}{2}\log\kappa^2 - \frac{\pi}{8}\kappa^2 \, {}_4F_3\left(1,1,\frac{3}{2},\frac{3}{2};2,2,2;\kappa^2\right)$$

$$I_1 + I_2 = \frac{\pi}{2}\log\left(\frac{\kappa^2}{4}\right) = -\pi\log 2 + \frac{\pi}{2}\log\kappa^2$$

$$I_1 = \frac{\pi}{2}\log 2 - \frac{\pi}{16}\kappa^2 \, {}_4F_3\left(1,1,\frac{3}{2},\frac{3}{2};2,2,2;\kappa^2\right)$$

The expression for $I_1$ can also be obtained using the differentiation under integral sign procedure [14].